\documentclass[aps,twocolumn,longbibliography,superscriptaddress]{revtex4-1}
\usepackage{epsfig}
\usepackage{epstopdf}
\usepackage{amsmath}
\usepackage{amsfonts}
\usepackage{amssymb}
\usepackage{hyperref}
\usepackage{bm}
\usepackage{makecell}
\usepackage{rotating}
\usepackage[utf8]{inputenc}
\usepackage{eurosym}

\usepackage{graphicx}
\usepackage{dcolumn}
\usepackage{bm}
\usepackage{color}

\usepackage{tikz,xcolor,hyperref}

\definecolor{lime}{HTML}{A6CE39}
\DeclareRobustCommand{\orcidicon}{%
	\begin{tikzpicture}
	\draw[lime, fill=lime] (0,0)
	circle [radius=0.16]
	node[white] {{\fontfamily{qag}\selectfont \tiny ID}};
	\draw[white, fill=white] (-0.0625,0.095)
	circle [radius=0.007];
	\end{tikzpicture}
	\hspace{-2mm}
}

\foreach \x in {A, ..., Z}{%
	\expandafter\xdef\csname orcid\x\endcsname{\noexpand\href{https://orcid.org/\csname orcidauthor\x\endcsname}{\noexpand\orcidicon}}
}

\begin{document}

\title{
From narrow-gap and semimagnetic semiconductors to spintronics and topological matter: A life with spins}


%



\author{Tomasz Dietl}
\affiliation{International Research Centre MagTop, Institute of Physics, Polish Academy of Sciences, Aleja Lotnikow 32/46, PL-02668 Warsaw, Poland}
\email{dietl@MagTop.ifpan.edu.pl}

\begin{abstract}
The abundance of semiconductors in our smartphones, computers, fiber optic junctions, cars, light sources, photovoltaic and thermoelectric cells results from the possibilities of controlling their properties through doping, lighting, and applying various fields. This paper, a part of the volume celebrating 100 years of the Polish Physical Society,  presents a biased selection of worthwhile results obtained by researchers at the Institute of Physics, Polish Academy of Sciences  relevant, as seen today, to topological matter and spintronics. Comprehensive studies, combining materials development, experimental investigations, and theoretical description of narrow-gap and dilute-magnetic semiconductors have been especially significant in this context. This survey also emphasizes , in an autobiographical tone,  a half of a century of author's intellectual emotions accompanying the rise of ideas and quantitative theories, allowing identifying the physics behind ongoing and future observations.
\end{abstract}

\maketitle
\tableofcontents

\section{Introduction}
In my recorded talk \cite{youtube}, I presented recent breakthroughs in spintronic and topological-matter research relevant to today's and tomorrow's information-communication technologies. Such a program was aimed at a broad audience of Polish Physical Society meetings -- teachers, publishers, colleagues from different fields. Not surprisingly, therefore, given the extraordinary and ceremonial character of the meeting, Darek Wasik, the session chair, and the semiconductor community fellow, commented the talk: "... as we are at the meeting of Polish physicists, I would like to underline their contributions and your Tomek to semimagnetic semiconductors and spintronics". Hence, this paper is an addendum to the presentation \cite{youtube}, and describes, in an ahistorical and self-centered  way, a couple of past accomplishments of researchers at the Institute of Physics, Polish Academy of Sciences (IFPAN) concerning narrow-gap and dilute magnetic semiconductors relevant, as seen today, to topological matter and spintronics.

In a sense, this survey is complementary to a broad overview of semiconductor research in Poland presented by Maria Kami\'nska from University of Warsaw's (UW) viewpoint \cite{Kaminska:2020_youtube}, and also to Jan Gaj contribution to a previous meeting of the Polish Physical Society \cite{Gaj:1994_PF} as well as to recollection reviews by Robert Ga{\l}{\c{a}}zka \cite{Galazka:2006_pssb} and Andrzej Kisiel {\em et al.}~\cite{Kisiel:2002}. Many systematized information can be found in a recent unsurpassed monumental masterpiece of Andrzej Wr\'oblewski \cite{Wroblewski:2020_B}. Obituaries and reminiscences recall  to us the importance of our Masters  and colleagues, Leonard Sosnowski (1911-1986) \cite{Grynberg:2002_PF}, Jerzy Mycielski (1930-1986) \cite{Blinowski:1986_PF,Furdyna:1986_PT}, Witold Giriat (1926-2001) \cite{Mycielski:2001_PF,Brazis:2015}, Jan Blinowski (1939–2002) \cite{Kacman:2002_PF,Turski:2002_PF}, Jan Gaj (1943-2011) \cite{Grynberg:2011_PF}, and  Micha{\l} Nawrocki (1943–2018) \cite{Suffczynski:2019_PF}. I have been unable to find a history of the Department  of Solid State Physics in Zabrze, specializing in II-V narrow-gap and dilute magnetic semiconductors, run for years by Lidia (1925-1996) and Witold (1923-2008) \.Zdanowicz (Gulag prisoner and Monte Casino hero). Also, there is no adequate account of the Unipress contribution to the physics of narrow-gap semiconductors, InSb in particular.

I hope that the present overview will trigger the appearance of papers showing the topic from different angles or describing developments of other physics branches in a similar way. This article purposely presents  only references to papers containing the IFPAN affiliation. One hundred reviews I am co-author of, particularly the two most recent \cite{Dietl:2014_RMP,Dietl:2015_RMP}, contain extensive and, hopefully, well-balanced citations of other institution contributions.


 %

\section{Surface and bulk Dirac cones in narrow-gap semiconductors: from a basement to an attic}
\label{sec:TCI}
I am not alone regarding the ARPES visualization of 2D topological surface Dirac states in Pb$_{1-x}$Sn$_x$Se \cite{Dziawa:2012_NM}, as IFPAN's most significant accomplishment in the second decade of the XXI century. These studies were initiated by Tomasz Story immediately after a suggestion of the MIT/Northeastern group that SnTe-type materials belong to a class of topological crystalline insulators (TCIs), just discovered theoretically by Liang Fu at MIT. The Warsaw/Lund/Stockholm results were not only submitted earlier than those of Sendai/Osaka and Princeton/Shanghai/Berkeley/Boston/Villigen/Huston/Tampei collaborations but showed the full gapless Dirac cones, not only their valence band part and, moreover, traced the topological phase transition in a single sample. Furthermore, Ryszard's Buczko tight-binding computations supported the photoemission data, and nicely illustrated the presence of surface gapless cones only if the band structure is inverted, i.e., anion $p$-states reside above cation ones.  Not surprisingly, therefore, this publication is already among ten most cited IFPAN's papers, and Tomasz Story has been invited to speak about at many venues, including the 32nd International Conference on the Physics of Semiconductors (ICPS; Austin TX 2014) and APS March Meeting (San Antonio TX 2015). For these studies, Andrzej Szczerbakow obtained high-quality Pb$_{1-x}$Sn$_x$Se single crystals by a 20\,k\euro \.self-selecting vapor growth method \cite{Szczerbakow:2005_PCGC}, perfected during his retainment times in a basement just below five pieces of ultrahigh vacuum equipment for MBE, sometimes refereed to as mega backs' epitaxy. Similarly to the case topological insulators of Bi and Sb chalcogenides developed at Princeton by Robert Cava's group, the IV-VI program at IFPAN had been carried on in the pre-topological epoch with thermoelectric applications in mind.

Experimental and theoretical studies of TCIs are continued in IFPAN on many fronts and also in a worldwide collaboration. Among others, Matthias Bode's STM group in W\"urzburg discovered, on surfaces of Szczerbakow's Pb$_{1-x}$Sn$_x$Se, 1D topological Dirac cones adjacent to odd surface step edges \cite{Sessi:2016_S}, presumably, the first observation of higher-order topological states. Present studies of topological surface bands, making used of wonderfully working Krak\'ow's SOLARIS ARPES facilities, include Weyl semimetals \cite{Grabecki:2020_PRB} and ferromagnetic TCIs \cite{Mazur:2019_PRB}, also examined from a theoretical perspective \cite{Brzezicki:2019_PRB,Lusakowski:2021_PRB}.

The case of TCIs, topological insulators, and earlier of graphene, shows that electron Dirac cones have been with material physicists for a long.  But when and where have they  been observed for the first time? This question brings us back to the beginning of the 1960s and to the attic of UW's physics building at Ho\.za 69. There, IFPAN's Witold Giriat initiated the grow of HgTe under the wings of Leonard Sosnowski, Master and  Mentor of many generations of Warsaw's semiconductor physicists, who, while in UK elucidated the physics of $p$-$n$ junctions (three Nature articles 1946-1947), and became elected as the President of the International Union of Pure and Applied Physics (IUPAP) for the term 1978-1981. From today's perspective, IFPAN studies of narrow-gap semiconductors brought several worthwhile accomplishments relevant to present works on topological matter of semiconductors and semimetals:

\ \\
(1) {\em Inverted band structure of HgTe}. Sylwester Porowski and co-workers developed at IFPAN high-pressure technique (the first step on Sylwester's heroic way to found at PAS the prominent Institute of High-Pressure Physics, Unipress). Measurements of the Hall resistance and the Seebeck coefficient $\alpha$ up 1.2\,GPa at room temperature demonstrated that HgTe has an inverted band structure \cite{Piotrzkowski:1965_pssb}, just proposed by Steven Groves and William Paul at Harvard for gray tin.  Robert Ga\l{\c{a}}zka (1937-2021), under supervision of Leonard Sosnowski, grew bulk n-Hg$_{0.9}$Cd$_{0.1}$Te and, by measuring $\alpha$ in high magnetic fields, determined the values of the effective mass $m^*$ of electrons as a function of their concentration $n$  \cite{Galazka:1967_pssb}. The results pointed to  $m^*\propto n^{1/3}$, proving a linear (Dirac) dispersion, $E(k) = v_fk$. Amazingly, the resulting Fermi velocity $v_f = 1.1\cdot10^8$\,cm/s is within 10\% of the graphene value. Despite the lack of topological protection, fine-tuning of the energy gap to the Dirac point by hydrostatic pressure made it possible to observe, a dozen years later, electron mobility as high as $2\cdot10^7$ cm$^2$/Vs in Hg$_{0.94}$Mn$_{0.06}$Te \cite{Sawicki:1983_Pr} (see Sec.\,\ref{sec:resonant}). The thermoelectric measurements of HgTe (and afterward of HgSe) were taken over and sophisticated by Andrzej J{\c{e}}drzejczak, my MSc supervisor (1972/1973), whose and my data I put to our first publications \cite{Dietl:1975_pssb,Jedrzejczak:1976_pssb}. Andrzej shared my wife Maniela and my happiness on our son Marek's birth, as his Godfather.

About that time, Andrzej Mycielski and Lucjan \'Sniadower studied optical properties of HgTe-based system, which brought Andrzej to find out with Jacek Baranowski (UW) infrared detecting capabilities of Hg$_{1-x}$Cd$_x$Te. Andrzej did not wait long to start supplying these detectors around, for instance, to mines as local heat sensors. J\'ozef Piotrowski took over and develop this technology in the Military University of Technology, and founded Vigo System, now a stock-exchange global company located in Warsaw's area, which produces top-level infrared detectors. This company is one of MagTop's industrial partners.

\ \\
(2) {\em Electron transport phenomena}. The quantitative description of above experimental data was carried out making use of Jerzy Ko{\l}odziejczak and Leonard Sosnowski's theory of transport phenomena in semiconductors with a non-parabolic band structure proposed for InSb by Evan Kane in his famous 1957 paper carried out at General Electric. One of the key concepts introduced by them that time was the energy-dependent momentum effective mass, $m^* = \hbar k/v = \hbar^2k(d\varepsilon/dk)^{-1}$. Leonard Sosnowski presented emerging IFPAN's experimental and theoretical results in his invited talks at the 7th and 9th ICPS (Paris 1964, Moscow 1968). Jerzy Ginter and Wanda Szyma\'nska were also members of the team, soon enriched by the dynamism of Wlodek Zawadzki. The crucial development, inspired by Henry Ehrenreich (1928-2008), then at General Electric and later at Harvard, was the use of proper Kohn-Luttinger amplitudes $u_{\bm{k}}$ that took properly into account $\bm{k}$-dependent spin-momentum locking \cite{Zawadzki:1971_pssb}. This formalism, summarized by Wlodek in his plenary talk at the 11th ICPS  (Warsaw 1972) and in comprehensive Adv.~Phys.~paper \cite{Zawadzki:1974_AP}, became a world standard in describing momentum, energy, and spin relaxation in narrow-gap semiconductors, InSb in particular. PhD students Piotr Bogus{\l}awski and me, in daily collaboration with Wanda Szyma\'nska, generalized this theory to zero-gap semiconductors \cite{Szymanska:1974_pssb,Szymanska:1978_JPCS}. That theory I converted to a comprehensive code which was successfully used for quantitative descriptions of my and others' electron mobility and thermomagnetic data taken over a wide temperature range and -- as we would say now -- across the topological phase transition in Hg$_{1-x}$Cd$_x$Se \cite{Dietl:1978_JPCS,Iwanowski:1978_JPCS,Iwanowski:1978_JPC} and Hg$_{1-x}$Cd$_x$Te \cite{Dubowski:1981_JPCS}.

\ \\
(3) {\em Relativistic effects and semirelativity}. As demonstrated in the last decade, topologically non-trivial states live at the boundaries of materials with inverted band structures, such as Hg-, Bi-, and Sn-based chalcogenides and cadmium arsenide. Relativistic effects, such as the spin-orbit interaction and the velocity-mass correction, stand behind the pertinent properties of topological matter, including spin-locking and the appearance of inverted band structures ($p$-type anion orbitals above  $s$ or even $p$ cation states) in materials containing heavy elements. This band inversion, quantified by topological invariances, results in gapless boundary states when stitching wave functions at interfaces between materials belonging to different topological classes.  At the same time, effective (multiband) Hamiltonians explaining quasi-particles properties in solids, particularly of electrons in narrow-gap semiconductors and semimetals, are often formally identical to various forms of equations put forward by Paul Dirac (1902-1984), Hermann Weyl (1885-1955), Ettore Majorana (1906-unknown), and Frank Wilczek to describe existing or hypothetical elementary particles, fermions in particular.

While the first report about an analogy between the four-band model (including spin) equations for narrow-gap semiconductors and the Dirac equation goes back to Leonid Keldysh (1931–2016), since 1964 Wlodek Zawadzki has persistently developed and extended this analogy \cite{Zawadzki:2017_JPC}, also in the reverse direction, for instance, by deriving a formula for the spin magnetic moment of relativistic electrons \cite{Zawadzki:1971_PRD}. In general, however, the four-band version of the Kohn-Luttinger $kp$ theory does not suffice to describe  semiconductor properties quantitatively. Accordingly, eight or fourteen band variants were developed at MIT, in Paris, IFPAN, and other labs in order to tackle the magneto-optical and magneto-transport phenomena in these systems.

\ \\
(4){\em Quantized magnetic fields}. Warsaw's physicists have often been taking their samples with and carried out charge transport and optical measurements in quantized magnetic fields in various labs around the globe. However, helium started to be available (Jerzy Rau{\l}uszkiewicz (1927-2005) took care of), and cryostats with superconducting coils were purchased in the beginning of the 1970s, when we moved to a new IFPAN campus on aleja Lotnik\'ow in 1973.  Andrzej Mycielski with a group of Ph.D. students, Margaret Dobrowolska among them, developed an infrared magnetooptical lab, while me with Daniel Dobrowolski and a gifted technician Piotr Siemi\'nski (1947-1984), a setup for measurements of the Shubnikov-de Haas (SdH) oscillations employing a field-modulation technique \cite{Dietl:1976_PAK}. Simultaneously, Daniel wrote a code providing Landau level energies within the Pigeon-Brown model, a key tool for the interpretation of magnetooptical and magnetotransport data. Our results, published in conference proceedings, have had a rather limited impact. That was also the fate of a report on the theoretical and experimental demonstration that a large difference between Dingle and mobility temperatures did not originate from samples inhomogeneities but could be accurately explained in terms of the quantum lifetime \cite{Dietl:1978_JdP}, the fact rediscovered at IBM a couple of years later. However, the built setups and codes, together with the already existing growth facilities, turned out to be essential for a quick take off with studies on semimagnetic semiconductors or, according to the terminology imposed in the US, dilute magnetic semiconductors (DMSs).

\section{The emergence of semimagnetic semiconductors}
\label{sec:DMS}
There are two events integrating semiconductor physicists in Poland: Friday seminars at UW, started by Leonard Sosnowski after his return from UK in 1948, and Jaszowiec annual meetings initiated by Witold Giriat, whose 12th and 49th editions were postponed because of the martial law and the pandemic situation in 1982 and 2020, respectively. During one of the first Friday seminars I attended, Robert Ga\l{\c{a}}zka, after his return from Jacek Furdyna's lab at Purdue, unfolded a vision of studying mainstream semiconductor compounds with group II or group III cations substituted partly by transition metal ions \cite{Galazka:1977_PF}. According to the painted picture, these new systems -- referred to as semimagnetic semiconductors -- should show some remarkable properties of magnetic semiconductors but without long-range order (hence the name semimagnetic). At the same time, owing to host's excellent semiconductor characteristics, they should be more accessible for investigating, understanding, and applying.

Within this program, Jacek Kossut at IFPAN, but under the supervision of Jerzy Mycielski at UW, began investigating theoretically electron scattering by magnetic impurities in narrow-gap semiconductors \cite{Kossut:1975_pssb,Kossut:1976_pssb}, Anna Paj{\c{a}}czkowska established that, typically, up to 50\% percent of Mn could be introduced to II-VI compounds   preserving excellent structural characteristics \cite{Pajaczkowska:1978_PCGC}, Andrzej Mycielski elaborated a method of Mn purification, and transferred the invention to his on-campus company (now PUREMAT Technologies, an industrial partner of MagTop) that has became the world leader in supplying pure Mn to research labs. Barbara Witkowska, an electrical engineer and then growth technology expert, has assisted Andrzej for half a century. Ursula D{\c{e}bska (1939-2020) (later at Purdue) and Andrzej Szczerbakow jointed the growth team. Though busy with writing of the Ph.D. thesis and associated papers \cite{Szymanska:1978_JPCS,Dietl:1978_JPCS,Iwanowski:1978_JPCS,Iwanowski:1978_JPC}, I was aware of the approaching wave, signalized by a striking behavior of SdH oscillations in Hg$_{1-x}$Mn$_x$Te \cite{Jaczynski:1978_pssb}, giant Faraday rotation in Cd$_{1-x}$Mn$_x$Te found at UW by Jan Gaj and Micha{\l} Nawrocki \cite{Gaj:1978_SSC} (soon commercialized in optical isolators by TOKIN company in the Sendai region), and Gerald Bastard's {\em et al.} magnetooptical data obtained for Hg$_{1-x}$Mn$_x$Te at ENS in Paris \cite{Bastard:1977_pssb,Bastard:1978_JdP}. I also attended Robert's invited talk at the 14th ICPS in Edinburgh, August 1978, in which those results made an extraordinary impact on the audience \cite{Galazka:1978_Pr}. At the same time (1977 and 1978) Kiev's group around Sergiy Ryabchenko reported giant exchange exciton splittings in CdTe:Mn and ZnTe:Mn, respectively.

After completing Ph.D., to build expertise, on Robert's recommendation, I spent 10 months in 1978 at an \'Ecole Polytechnique lab founded by Ionel Solomon (1929-2015), which in collaboration and competition with physicists around Boris Zakharchenya (1928-2005) at Ioffe, made pioneering contributions to spintronics associated with spin-orbit and hyperfine interactions in semiconductors. Despite a decade elapsed, May 1968 was in the air. Georges Lampel, a discover of spin pumping in semiconductors, as a protest against the incarceration of physicist Yuri Orlov and  mathematician Natan Sharansky  declined to be an editor of the reference book {\em Optical Orientation} (North Holland 1984) in the series containing chapters written alternately by Western and Soviet physicists. At the same time, \'Ecole Polytechnique elite students demanded supplying {\em L'Humanit\'e} (indirectly supported by Soviets) to the Palaiseau campus newspaper stands. On arrival, Ionel -- my Master --  showed me our working horse, an automatized setup for resistivity measurements of hydrogenated amorphous silicon, equipped with Commodore, the first personal computer I had ever seen. Nevertheless, we all participated in a 45\,min protest (making me a strike expert in the Solidarity carnival times)  when a secretary did not obtain an increase in salary after refusing to attend a computer course.

By field-effect measurements, we showed that surface depletion layers control the film properties \cite{Solomon:1978_JdP}. I think my main accomplishment was to invalidate, by a series of experiments, Ionel's idea that the Staebler–Wronski effect, killing photovoltaic cells' performance, is a surface phenomenon. Amorphous silicon expertise introduced me to the physics of disordered  (Sec.~\ref{sec:MIT}) and 2D systems (see, Sec.~\ref{sec:structures}). At the same time, reading for a change older and newer local Ph.D. theses, I became familiar with the fascinating world of spin-dependent phenomena in solids. On Paris diners, with my distant cousins Andrzej and Jurek Mycielski (visiting Claudette Rigaux's lab) as well as with G\'erald Bastard, Yves Guldner, and Jan Gaj [staying in the Claude Benoit \`a la Guillaume (1925-1994) group], the bright prospects of DMSs became clear to me.

On arrival to Warsaw, with my MSc student Marcin Otto, I built the first Warsaw's setup for magnetization measurements of DMSs, an a.c. susceptometer. We did not find any effects of electron concentration changes by annealing in Hg$_{1-x}$Mn$_x$Se,  but provided supplementary information for Margaret's magnetooptical data \cite{Dobrowolska:1980_JPSJ} and, in parallel to Robert's SQUID measurements at Purdue, detected spin-glass freezing in Hg$_{1-x}$Mn$_x$Te, the results published somewhat later with Faraday rotation data \cite{Mycielski:1984_SSC}. The magnetization behavior in n-Cd$_{1-x}$Mn$_x$Se was essential in building up the understanding of the physics of bound magnetic polarons \cite{Dietl:1982_PRL}. To learn more about how to change carrier density by the field-effect, I spent July 1980 in Mecca of 2D semiconductor systems -- Fred Koch (1937-2012) lab at the Technical University of Munich, where I heard Klaus von Klitzing announcing, perhaps for the first time, his discovery of the quantum Hall effect \cite{Dietl:1982_PF}. I also spoke there about DMSs, and we both were recommended by Fred to Noboru Miura as invited speakers at the Oji International Seminar on High Magnetic Fields in Semiconductor Physics in Hakone, September 1980, a satellite event to the 15th ICPS in Kyoto, where Jan Gaj also delivered an invited talk on DMSs. With Fujiyama in the window, I presented an overview of Warsaw/Paris findings, adopting the $s$-$d$ Vonsovskii model to (II,Mn)VI compounds and pointing out that $p$-$d$ hybridization accounts for exchange coupling of $\Gamma_8$ carriers to Mn spins, the suggestion confirmed theoretically by Anadi Bhattacharjee {\em et al.} in Orsay.

I would single out five durable accomplishments of studies devoted to bulk DMSs, partly described in early reviews \cite{Furdyna:1988_B,Dietl:1994_HB}:

\ \\
(1) {\em $sp$-$d$ exchange energies}. To interpret his inter-band magnetooptics data on n-Hg$_{1-x}$Mn$_x$Te, Gerald Bastard with a support of Jerzy Mycielski described the influence of Mn ions on the effective mass electrons in terms of the molecular-field and virtual-crystal approximations \cite{Bastard:1977_pssb,Bastard:1978_JdP}, an approach taken over by Jacek Kossut to interpret Marek Jaczy\'nski's SdH results \cite{Jaczynski:1978_pssb} and by Gaj, Ginter, Ga{{\l}}\c{a}}zka (the 3G model), who assumed that to quantify giant $sp$-$d$ exchange splitting of free exciton states, the Landau quantization can be neglected in wide-band gap semiconductors \cite{Gaj:1978_pssb}. These models, with minor modifications, have served for decades to describe magnetooptical and magnetotransport results and, thus, to determine the $s$-$d$ and $p$-$d$ exchange integrals (denoted as $\alpha$ and $\beta$, respectively) in a dozen of II-VI compounds containing cation-substitutional Mn ions but also Cr, Fe, and Co ions, the activity carried on by the Warsaw/Paris collaboration and, somewhat later, primarily by Andrzej Twardowski and co-workers at UW, Jacek Furdyna's group  at Purdue/Nore Dame, Don Heiman's {\em et al.} at MIT, Shojiro Takeyama and co-workers in Himeji and Tokyo, and other teams.  {\em Ab initio} computations by Henry Ehrenreich group at Harvard and Alex Zunger {\em et al.} in Golden CO as well as a tight-binding approach put forward by Jan Blinowski (UW) and Per{{\l}}a Kacman (IFPAN) allowed us to understand the signs and magnitudes of $\alpha$ and $\beta$ \cite{Kacman:2001_SST}. A similar approach \cite{Dietl:1994_PRB} has been, at least semiquantitatively, successful in describing  magnetooptical splittings in PbTe and PbSe with substitutional Mn and Eu cations \cite{Bauer:1992_SST} and the EPR Knight shift in the of Mn in PbTe and SnTe \cite{Story:1996_PRL}. We continue this line of research extending the theory for arbitrary ${\bm{k}}$ \cite{Autieri:2021_PRB}.

\ \\
(2) {\em Strong coupling effects}. Attempts to go beyond virtual-crystal and molecular-field approximations have a long history, often put forward in the context of the so-called mismatched alloys. The key insight, which I share, is that, similarly to the Kondo effect and superconductivity, perturbation approaches come short, as a substantial modification of the local potential means that a bound state can be formed, even for the substitution by the element with the same valence, such as N in GaAs, Mn in ZnO, or Fe in GaN. A non-perturbative Wigner-Seitz-like approach to the $p$-$d$ exchange interaction that grows up with decreasing the anion-cation bond length was proposed by Claude Benoit \`a la Guillaume and Denis Scalbert, to which I incorporated the non-magnetic alloy potential \cite{Benoit:1992_PRB}. Our work explained a mysterious increase of the exchange energy $|N_0\beta|$ with lowering $x$ observed by exciton magnetospectroscopy for Cd$_{1-x}$Mn$_x$S.  In a series of beautiful works Jakub Tworzyd{\l}o at UW sum up a class of diagrams describing self-energy of holes at the top of the valence band in the presence of randomly distributed Mn spins of arbitrary magnetization, and demonstrated the quantitative accuracy of our model for Cd$_{1-x}$Mn$_x$S.  However, even more surprising was outcome of magnetooptical studies carried out for ZnO and GaN doped with Co and Mn in the middle of the 2000s by Wojciech Pacuski and co-workers at UW, Grenoble, and IFPAN. The results implied the $N_0\beta$ values of  opposite sign and much reduced amplitude compared to those stemming from photoemission and XAS as well as expected from the chemical trends. I realized that  Jakub's approach can be extended to the case, when the hole is bound to the transition metal ion, and showed that the model explains the reversal of the valence band splitting \cite{Dietl:2008_PRB}, in agreement with experimental data for Ga$_{1-x}$Fe$_x$N \cite{Pacuski:2008_PRL} and  Zn$_{1-x}$Mn$_x$O \cite{Pacuski:2011_PRB}.  I think that splitting of the conduction band into two branches observed in mismatch alloys could be explained by Jakub's theory generalized for $k \ne 0$.

Somewhat related is the issue of atypical magnitudes of $\alpha$ reported for certain DMSs. For instance, David Awschalom's group found negative values of $\alpha$ for paramagnetic Ga$_{1-x}$Mn$_x$As with $x \le 0.13$\%, which I assigned to the exchange interaction with the hole residing on the Mn acceptor, the suggestion wonderfully quantified by Czarek \'Sliwa \cite{Sliwa:2008_PRB}.

\ \\
(3) {\em Bound magnetic polarons}. Do effective electrons affect Mn spins? Around 1980 bound magnetic polarons (BMPs) have been uncovered optically by the Paris/Warsaw collaboration. Particularly clear were  Micha{{\l}} Nawrocki {\em et al.} spin-flip Raman scattering (SFRS) data for n-Cd$_{1-x}$Mn$_x$Se \cite{Nawrocki:1981_PRL}, which revealed spin splitting of donor electrons even in the absence of Mn macroscopic magnetization. In 1980/1981 J\'ozek Spa{\l}ek started coming from Krak\'ow and brought to us his knowledge about magnetic semiconductors, particularly on theories of BMPs developed for those materials. It became clear to me that previous approaches missed the key ingredient -- the significance of thermodynamic magnetization fluctuations in the case of localized electrons. To resolve the central-spin problem, as we would say now, I learned the Ginzburg-Landau approach to phase transitions  from the book of Shang-keng Ma (1940-1983), and formulated the question in terms of functional integrals without the "box" approximation \cite{Dietl:1983_PRB,Dietl:1983_JMMM}. With J\'ozek, we nicely interpreted Micha{{\l}}'s SFRS and my magnetization results \cite{Dietl:1982_PRL}. Here, we were faster than the MIT group around Peter Wolff (1923–2013), and soon the Dietl-Spa{{\l}}ek model, yielding the shift, width, and shape of the spin-flip line has been verified by a dozen of groups, also in the context of widely studied excitonic magnetic polarons (EMPs). Later Tomasz Wojtowicz, using Witold Plesiewicz's homemade SQUID, confirmed quantitatively our predictions concerning BMP magnetization \cite{Wojtowicz:1993_PRL}.

Works on BMPs made that I have been internationally recognized  as a theoretician. In that hat, I interpreted with the G\"unther Bauer group photomagnetization data \cite{Krenn:1989_PRB} and later, while in Grenoble, results on the formation time of EMPs obtained by my hosts and  previously at Brown, MIT, and IBM \cite{Dietl:1995_PRL}. At the same time, I was rightly sceptical about the idea of the free magnetic polaron appearing in the DMS literature at that time. More recently, in 2015, I confirmed that the central spin problem can be solved semiclassically, i.e., I reproduced accurately within our formalism the recent results obtained by solving the quantum Liouville equation for the whole system or from quantum Monte Carlo numerical simulations, concerning the dynamics of the localized electron spin in the hyperfine field of nuclear magnetic moments \cite{Dietl:2015_PRB}.

\ \\
(4) {\em Antiferromagnetic superexchange}.
There is a consensus that the exchange interaction between dilute Mn ions in II-VI and IV-VI semiconductors is bilinear in spin operators and antiferromagnetic for all distances between magnetic ions \cite{Kacman:2001_SST}. The Hamiltonian contains the scalar Heisenberg and the pseudo-dipole terms and, depending on the spin pair symmetry in a given host, non-scalar contributions, such as the Dzyaloshinskii-Moriya component, visible clearly in EPR studies. As suggested by J\'ozek Spa{\l}ek {\em et al.} \cite{Spalek:1986_PRB}, and shown quantitatively by the Ehrenreich group at Harvard, the short-range superexchange is the dominant interaction mechanism in Mn-based II-VI DMSs. A linear dependence of the Curie-Weiss temperature on the Curie constant in the high temperature limit \cite{Spalek:1986_PRB} proved the uncorrelated distribution of Mn ions and also Co ions \cite{Sawicki:2013_PRB}. Inelastic neutron and light scattering, together with steps in $M(H)$ dependencies in high magnetic fields, have provided quantitative information on exchange energies for Mn, Co, and Eu nearest-neighbor pairs in II-VI and IV-VI DMSs \cite{Bonanni:2021_HB}. Low-temperature specific heat and magnetization measurements carried out in various labs across the globe allowed establishing that freezing temperature $T_{\text{f}}  \propto x^p$, where $p = 1.33$ for (Cd$_{1-x}$Mn$_x$)$_3$As$_2$ and $p = 2.3$ for wide band-gap Mn-based DMSs \cite{Galazka:2006_pssb}.  Faraday rotation \cite{Leclercq:1993_PRB} and quantum noise \cite{Jaroszynski:1998_PRL} served to determine the character of spin-glass dynamics.

\ \\
(5) {\em Playing with magnetism in magnetically doped Pb$_{1-x}$Sn$_{x}$Te}.
Two pioneering works of Tomasz Story {\em et al.}  demonstrated that  it is possible, by changing carrier density, to trigger {\em ferromagnetism} in Pb$_{1-x-y}$Sn$_y$Mn$_x$Te \cite{Story:1986_PRL} and to alter the strength of {\em antiferromagnetic} coupling in Sn$_{1-x}$Gd$_x$Te \cite{Story:1996_PRLa}. In the former case, the intrinsic antiferromagnetic interaction of Mn ions is overcompensated by carrier-mediated RKKY-type ferromagnetic coupling once the holes start to occupy twelve side $\Sigma$ valleys, as revealed by magnetization studies under hydrostatic pressure \cite{Suski:1987_JMMM}.  In the case of Sn$_{1-x}$Gd$_x$Te, Gd ions introduce occupied donor and empty acceptor states  (5$d^1$ and 5$d^2$, respectively), which are resonant with the valence band.  By annealing, one decreases the concentration of native acceptors, which shifts the Fermi energy toward the $5d^2$ level. According to the data and the model put forward in Ref.~\cite{Story:1996_PRLa}, antiferromagnetic coupling between Gd spins becomes then resonantly enhanced. In those systems vacancy-related hole densities attain the level of $7\cdot10^{20}$\,cm$^{-3}$, not reachable in the case of II-VI DMSs but available in Ga$_{1-x}$Mn$_x$As, where Mn ions act as acceptors providing holes to a relatively simple and well-known valence band (see, Sec.~\ref{sec:saga}). As already mentioned (see, Sec.~\ref{sec:TCI}), those materials combine magnetism with topological characteristics, and are in the center of MagTop studies.

\section{Why do spin effects account for positive magnetoresistance, whereas orbital phenomena for negative MR, and not {\em vice versa}?}
\label{sec:MIT}
"What about spin?" I asked Phil Anderson (1923-2020) to break the embarrassing silence after his plenary talk at Montpellier's IUPAP 16th ICPS in 1982 \cite{Dietl:1983_PF}. In his lecture, he presented the Gang of Four work on quantum localization [inspired by David Thouless (1934-2019)] and corresponding millikelvin results taken at Bell Labs. He jumped on the topic as, according to his later writings, he spent last evenings discussing to what extent  spin physics could account for the apparent disagreement between the theory and data on the quantum metal-to-insulator transition (MIT) in Si:P. My perspective was different: in IFPAN, we just began millikelvin measurements of Hg$_{1-x}$Mn$_x$Te \cite{Sawicki:1983_Pr} (see Sec.~\ref{sec:resonant}) and magnetoresistance studies of  Cd$_{1-x}$Mn$_x$Se \cite{Dietl:1983_Pr} and p-Hg$_{1-x}$Mn$_x$Te \cite{Wojtowicz:1983_Pr} in the vicinity of the MIT.

The textbook paradigm was that the Lorentz force results in a positive magnetoresistance (MR), whereas spin-disorder scattering (especially in the Kondo limit) and bound magnetic polarons make MR negative in dilute magnetic materials and magnetic semiconductors. In the case of n-Cd$_{1-x}$Mn$_x$Se it was apparently another way around: the weak-field MR was negative in n-CdSe and positive in paramagnetic n-Cd$_{0.95}$Mn$_{0.05}$Se \cite{Dietl:1983_Pr}! Immersed in a summer scenery of southern France, so bright after a long martial-law darkness in Soviet-controlled Poland, I realized that DMSs (see Sec.~\ref{sec:DMS}) once more opened new research horizons -- this time to test fundamentally new theoretical predictions concerning the crucial role of subtle interference effects in one-electron and many-body Anderson-Mott localization in disordered systems, put forward by Boris Altshuler, Arkadii Aronov (1939-1994), Hidetoshi Fukuyama, Dima Khmelnitskii, Patrick Lee, T.V. Ramakrishnan, and others, the understanding accelerated by a Soviet-US joint meeting at Armenian's Lake Sevan in September 1979. When I learned that Jacek Kossut would host Yoshi Ono from the University of Tokyo, I asked them to consider MR of DMSs caused by quantum localization. Their paper considering the one-electron case \cite{Ono:1984_JPSJ} was a good starting point to develop a more comprehensive numerical code, the task taken over by Maciej Sawicki.

Importantly, we had in place already that time bulk DMS growth technology (developed first under the leadership of Witold Giriat and then of Robert Ga{{\l}}{\c{a}zka and Andrzej Mycielski, see Sec.\,\ref{sec:DMS})), and a dilution fridge program (initiated in IFPAN by Lucjan \'Sniadower and Piotr S{\c{e}}kowski, who then emigrated to France and Germany, respectively), and taken over in 1982 by a highly talented self-made cryogenic expert Witold Plesiewicz and me. Witek, till 2010 fabricated about 70 helium and nitrogen cryostats, distributed over whole Poland, but also exported to, e.g., University of Tokyo and Z\"urich's ETH.  He was my wedding witness, we were used to do not miss any illegal Solidarity demonstration (dressed in running shoes), and sequentially hosted in our apartments the Editorial Office of the underground Jan Str{\c{e}kowski's "Tygodnik Wojenny" ({\em War Weekly}) appearing in the years 1982-1985 and later the recording studio of the Solidarity Program II radio. Actually, our most active IFPAN colleague and future Senate member Zbyszek Romaszewski was sentenced for four and a half years of prison for organizing the first Solidarity radio. Zbyszek and his wife Zosia were ones of those who brought to falling walls in 1989.  His Moscow visit as an IFPAN delegate was in fact a cover for the famous meeting between KOR and Andrei Sakharov in 1979.

There were four new accomplishments collected in Ph.D. theses of Tomasz Wojtowicz (1988), Maciej Sawicki (1990), Jan Jaroszy\'nski (1990), and Pawe{\l} G{\l}\'od (1995), made possible by their contribution to the development of millikelvin measurement setups:

\ \\
(1){\em Effect of spin splitting}.
The giant $s$-$d$ exchange splitting produces a field-dependent mass in the diffusion and Cooperon poles in quantum conductivity corrections brought about by electron-electron interactions, which leads to a sizable {\em positive} MR on the conducting side of the MIT. The effect was initially found and interpreted quantitatively in n-Cd$_{1-x}$Mn$_x$Se:In \cite{Sawicki:1986_PRL}, and later investigated in various DMSs around the world, but also in IFPAN on epilayers from multiple labs,  n-Zn$_{1-x}$Mn$_x$O:Al \cite{Andrearczyk:2005_PRB}, n-Zn$_{1-x}$Co$_x$O:Al \cite{Dietl:2007_PRB}, Cd$_{1-x}$Mn$_x$Te HEMTs \cite{Jaroszynski:2007_PRB}, and n-Ga$_{1-x}$Mn$_x$N:Si \cite{Adhikari:2015_PRB}. Simultaneously, somewhat surprisingly to many, effects of spin-disorder scattering upon many-body quantum localization are not relevant in paramagnetic DMSs, as temperature at which thermal broadening $k_{\text{B}}T$ becomes smaller than spin-disorder scattering rate corresponds to the onset of carrier-driven ferromagnetic ordering of localized spins.

\ \\
(2) {\em Temperature-dependent localization}. We discovered an abrupt conductivity drop below 1\,K accompanying by a negative colossal MR (CMR) in the MIT's close vicinity in n-Cd$_{1-x}$Mn$_x$Se \cite{Sawicki:1986_PRL,Dietl:1986_PS,Glod:1994_PB}. To make a long story short: the key notion here is the disorder-driven spatial separation into regions with larger and smaller carrier density near the MIT, which -- owing to the carrier-mediated ferromagnetic coupling between Mn spins -- results in mesoscopic magnetization fluctuations leading to quantum localization \cite{Jaroszynski:2007_PRB,Dietl:2008_JPSJ}. Thus, the physics behind the phenomenon is similar to that put forward by Elbio Dagotto, Adriana Moreo and others as well as  Eduard Nagaev (1934-2002) to explain the origin of CMR in ferromagnetic oxides and semiconductors in the vicinity of $T_{\text{C}}$. Thus, within this model, critical carrier scattering and the associated negative CMR originate rather from spatial fluctuations of $T_{\text{C}}$ in the MIT neighborhood than from thermodynamic magnetization fluctuations at $T_{\text{C}}$ (which is below 100\,mK in the studied n-type DMSs).

\ \\
(3} {\em Critical behavior}. It was realized by Steve von Moln\`ar at IBM and by us that while quantitative modeling of conductance and, in particular, of negative CMR is not yet possible across the MIT, the presence of the field-induced insulator-to-metal transition in appropriately selected samples allows one to determine critical exponents of the Anderson-Mott quantum transition, a much-debated issue that time.  Especially attractive was the case of p-Hg$_{1-x}$Mn$_x$Te, in which, as shown by Jerzy Mycielski, an increase of $p$-$d$ exchange splitting makes the localization length to be more and more determined by the $\Gamma_8$ light hole mass. This effect strongly enhances CMR, and  allowed us to study the critical behavior of resistance, resistance anisotropy,  Hall effect, dielectric constant, and hopping length \cite{Wojtowicz:1986_PRL,Wojtowicz:1989_PB,Jaroszynski:1992_PB}. I called the Landau Institute and asked Sasha Finkelstein to provide us with dynamic renormalization group equations for the spin-polarized case, which we compared to experimental data \cite{Wojtowicz:1986_PRL}, the procedure later followed by many groups for the MIT in various disordered systems.

\ \\
(4) {\em Rashba weak antilocalization}. We were the first to reveal the influence of the Rashba spin-orbit term on transport phenomena experimentally by observing a tinny positive MR (weak antilocalization -- WAL) in n-CdSe:In below 1~K, which was taken over above 20~\,mT by a stronger negative MR -- a manifestation of the Aharonov-Bohm-type of interferences for electrons diffusing in disordered media (weak localization MR) \cite{Sawicki:1986_PRL}. Notably, the determined magnitude of the Rashba parameter was in excellent agreement with the value known from the strength of electron spin-resonance in n-Cd$_{1-x}$Mn$_x$Se, studied experimentally by Jacek Furdyna's and Denis Drew's groups, and theoretically by Peter Wolff at MIT.  The WAL MR was even smaller, but detectable, in n-ZnO:Al \cite{Andrearczyk:2005_PRB} and n-GaN:Si \cite{Stefanowicz:2014_PRB}. At the same time, our data allowed verifying, presumably for the first time quantitatively, the theory of the phase coherence time, confirming the dominant role of carrier-carrier scattering at low temperatures \cite{Sawicki:1986_PRL}.
\ \\

The innovative and comprehensive nature of this research resulted in inviting me to speak at IUPAP's 18th International Conference on Low Temperature Physics (Kyoto 1987) \cite{Dietl:1987_JJAP} and 19th ICPS, Warsaw 1988 \cite{Dietl:1988_Pr} as well as at the Meetings of the Condensed Matter Division of EPS (Budapest 1988) and of APS (St. Louis MO 1989).
That time, we carried out with Don Heiman millikelvin spin-flip Raman scattering measurements at the MIT Francis Bitter Magnet Lab, unsuccessfully looking for BMPs on the metal side of MIT in  n-Cd$_{1-x}$Mn$_x$Se \cite{Dietl:1991_PRB}. On those occasions, and with the help of Steve von Moln\'ar (1935–2020), I made two seminar tourn\'ees over a dozen US universities and research centers (including Bell Labs and IBM) traveling between them on Greyhound busses. Quite a progress compared to my lowest-level construction jobs in London and Brussels suburbs making possible seven hitchhiking summer tours between youth hostels and my beloved relatives homes distributed over magnificent West and South Europe cities (1967-1974; 25\,000\,km).

For the 40th birthday, I obtained a professor title in 1990. I was also asked to chair the International Conference on Electron Localization and Quantum Transport. The conference was organized in Jaszowiec with the logistic help of Ela Zipper at Silesian University but the banquet we arranged near Krak\'ow in a spacious grotto 100\,m below the ground, a part of the Middle Age Wieliczka Salt Mine complex -- the UNESCO World Heritage Site. One of the scheduled highlights was the after-dinner talk to be given by Nevill Mott (1905-1996) \cite{Dietl:1997_PF} about his Goettingen times with Max Born. Sir Nevill passed away that night, canceling a couple of days earlier his trip to Poland because of a flu infection.

There have been many interesting research followups. For instance, I found that surprising {\em negative} MR of ferromagnetic p-Ga$_{1-x}$Mn$_x$As persisting  up to at least 27\,T  at $ T \ll T_{\text{C}}$ \cite{Omiya:2000_PE} can be quantitatively interpreted in terms of the {\em orbital} single-electron WL interference effect \cite{Dietl:2003_Pr,Matsukura:2004_PE}, as the exchange band splitting is field-independent at low temperatures and also precludes WAL's MR appearance in ferromagnets. This insight was confirmed by comprehensive MR studies of p-Ga$_{1-x}$Mn$_x$As carried out by Dieter Weiss's group in Regensburg, the weak-field data showing, additionally,  the presence of demagnetization effects and field-dependent magnon scattering at non-zero temperatures. Another line of recent studies concerns ferromagnetic and nonmagnetic topological semiconductors which exhibit negative WL MR and positive WAL MR, respectively (see, e.g.~Ref.\,\onlinecite{Kazakov:2021_PRB}).

Finally, I have to note that despite all those developments there have been systematic assignments of negative MR and/or non-linearities in the Hall resistance to the chiral anomaly or to spin-disorder scattering by the existing magnetic impurities or by hypothetical spins (or two-level systems) on defects, prior to checking the role of orbital WL MR, whose magnitude can be evaluated without any adjustable parameters. Another comment concerns positive MR -- it obviously occurs in the case of multivalley or multilayer transport. Here, the application of the so-called mobility spectrum method can serve to determine densities and mobilities of contributing carriers. Furthermore, inhomogeneities-induced mixing between $xx$ and $xy$ resistivity tensor components results in a linear high-field positive MR.

\section{How do resonant states enhance carrier mobility?}
\label{sec:resonant}
The question on whether and under which conditions localized and extended states can co-exist at the same energy appears in many contexts, and is also central to the condensed matter physics, as it concerns, for instance, the survival of magnetic moments in metals and the issue of disorder-driven metal-insulator transition. The key notion is resonant scattering that dramatically lowers conductivity of metals containing impurities if they give rise to quasi-localized states with binding energies near the Fermi energy. Not surprisingly, this question was addressed for acceptors in zero-gap semiconductors, such as HgTe, for which Boris Gelmont and Mikhail Dyakonov  predicted in 1972 a greater binding energy than lifetime broadening. This created an abundance of theories providing the strength of electron resonant scattering, a prediction apparently confirmed by a minimum in the HgTe conductance at temperatures at which the Fermi energy assumes the expected positions of acceptors states, always present in real materials.

However, this insight was called into question by Wladek Walukiewicz who showed in a qualitative model that optical phonon scattering between the conduction and valence bands might account for the electron conductivity minimum \cite{Walukiewicz:1976_JPC}. Our comprehensive eight band $kp$ theory of electron transport in zero- and narrow-gap zinc-blende semiconductors, developed with Wanda Szyma\'nska employing a variational solution of the Boltzmann equation \cite{Szymanska:1978_JPCS} and verified for HgSe \cite{Dietl:1978_JPCS} and Hg$_{1-x}$Cd$_x$Se \cite{Iwanowski:1978_JPCS}, confirmed Wladek's suggestion by showing good agreement of the theory and experimental data for Hg$_{1-x}$Cd$_x$Te as a function of temperature and $x$ with no adjustable parameters \cite{Dietl:1978_Pr,Dubowski:1981_JPCS}.

So, are resonant states irrelevant in semiconductors? Absolutely not -- they can lead to a significant ... {\em enhancement} of carrier mobility. To describe magnetic impurities in metals, Anderson considered a competition between {\em intra}-site correlation energy $U$ and hybridization between impurity and band state $V_{\bm{k}d}$. In the case of resonant states in semiconductors, I argued \cite{Dietl:1987_JJAP,Wilamowski:1990_SSC}, that a competition of $V_{\bm{k}d}$ with {\em inter}-site Coulomb energy $E_{\text{C}}$ is essential. If the width of the Efros-Shklovskii gap produced by $E_{\text{C}}$ is larger than the tunneling rate from the resonant state to the band, the efficiency of resonant scattering will be much reduced. Moreover, as noted already in 1983 by the Nikolaj Brandt's (1923-2015) group at Moscow University, who investigated Witold Giriat's zero-gap donor-compensated p-Hg$_{1-x}$Mn$_x$Te and by Jerzy Mycielski in the context of Andrzej Mycielski's HgSe:Fe \cite{Mycielski:1988_JAP}, $E_{\text{C}}$ results in a correlated spatial arrangement of charges on resonant impurities, which dramatically enhances electron mobility and reduces the Dingle broadening of Landau levels at low temperatures. This model (i) explains quantitatively mobility magnitudes in HgSe:Fe in which Fe$^{3+}$/Fe$^{2+}$ donor states reside 0.2\,eV above the conduction band bottom \cite{Wilamowski:1990_SSC,Kossut:1990_SST}; (ii) elucidates the mechanism leading to mobility as high as $20\cdot10^{6}$\,cm$^2$/Vs found in semimetalic p-Hg$_{0.94}$Mn$_{0.06}$Te fine-tuned by hydrostatic pressure to the Dirac cone band arrangement \cite{Sawicki:1983_Pr}, and (iii) explains the pressure dependence of electron mobility in n-GaAs, in which Si donors form DX resonant states \cite{Suski:1990_SST}.

Jacek Kossut was invited to present IFPAN results at the 20th ICPS, Thessaloniki 1990. One evening of the meeting, I participated in the chair dinner in an elegant beach hotel. It was exactly those dining illuminated tables I observed {\em via} a fence, with jealousy in my eyes, one year earlier eating just cooked noodles in front of our little tent. (Family dinner in that restaurant would then cost my IFPAN monthly salary even though our 10 years old daughter Zosia gave up this sightseeing tour, preferring to take advantage of the last chance to visit East Germany with her dancing group.)

It was realized somewhat later that under modulation doping conditions, the barrier dopants constitute resonant states for the channel carriers. Amazingly, also in that case, ordering of impurity charges accounts for high electron mobility values \cite{Suski:1994_PRB}, reaching now $35\cdot10^{6}$\,cm$^2$/Vs in Si modulation-doped GaAs HEMTs grown by Loren Pfeiffer, Vladimir Umansky, Werner Wegsheider, and others, if a background ionized impurity concentration is reduced by efficient pumping and high purity of elements employed for MBE. More recently, I extended the charge ordering model for MBE-grown HgTe quantum wells (QWs) in the vicinity of the topological phase transition \cite{Dietl:2023_PRL,Dietl:2023_PRB}. In that case, I showed, residual acceptors form a resonant impurity band degenerate with the hole part of the QW Dirac cone. In this way I explained puzzling results obtained by Laurens Molenkamp's group and us \cite{Yahniuk:2021_arXiv}, in particular, why compared to electrons: (i) the hole mobility is large and the corresponding Dingle temperature low;  (ii) the hole concentration increases slowly with the gate voltage, and (iii)  the ground-state quantum Hall plateau of holes is exceptionally wide.

Interestingly but counterintuitively, the carrier mobility appears to be larger at the topological phase transitions in (Hg,Mn)Te compared to (Hg,Cd)Te. I have recently argued \cite{Dietl:2023_PRL,Dietl:2023_PRB} that the formation of BMPs around resonant acceptor states, by hardening of the Coulomb gap and diminishing Kondo scattering by acceptor holes, enhances the magnitudes of hole and electron mobility in 2D and 3D (Hg,Mn)Te samples, respectively.

\section{Quantum structures of dilute magnetic semiconductors}
\label{sec:structures}

Shortly after breaking out into independence, Poland joined CERN in 1991.
Other initiatives got a bust, too. In 1992, Robert Ga{\l}{\c{a}}zka and Jacek Kossut, IFPAN's Director and
Deputy Director, respectively, initiated a new MBE program with Jacek Kossut and
Tomasz Wojtowicz as persons in charge. In 1993 they made operating the new
MBE Lab equipped with the first commercial MBE system in Poland, purchased
from the EPI company in the USA, using IFPAN own funds. Simultaneously,
government subsidies financed a  couple of research groups interested in
low-dimensional systems. Jan Gaj coordinated the project, and Manijeh
Razeghi, a pioneer in epitaxial techniques for semiconductors, chaired a
panel of external advisors.  This initiative made it possible to develop, with
significant involvement of Grzegorz Karczewski and El{\.z}bieta Janik,
the MBE growth technology of II-VI compounds. At the same time, Jerzy Wr\'obel and I, in collaboration with  Eliana Kami\'nska and Anna Piotrowska at the Institute of Electron Technology,  constructed a clean room at IFPAN equipped with electron beam lithography and auxiliary nanofabrication tools. They were funded by other grants, including the Austrian Ost-West program we applied to with G\"unther Bauer with whom many other IFPAN colleagues and I had a long record of friendly and efficient collaboration.  Jerzy's proficiency is best illustrated by his Stern-Gerlach solid-state spin-filter, whose fabrication required five electronolithography levels and lift-off of four different metals \cite{Wrobel:2004_PRL}. It looks, he triggered the use of micromagnets by the spin-qubit community.  IFPAN's epitaxy and nanofabrication capabilities have been enriched by Tomasz Wojciechowski and Tomasz Wojtowicz's decades-long tireless activities, culminating by the opening in 2017 of the new Laboratory of Technological Processes of Semiconductor Nanostructures and Devices, now further expanding within the MagTop project.

Of course, the acquiring of funds for the quantum structure programme was facilitated by our earlier accomplishments in the physics of low-dimensional systems \cite{Dietl:1986_APPA,Dietl:1993_SST}. For instance, my then Ph.D. student Grzegorz Grabecki had started in 1980 fabricating field-effect transistors of p-Hg$_{1-x}$Mn$_x$Te, and successfully demonstrated the influence of the $sp$-$d$ exchange interaction on the SdH oscillations of interfacial electrons as a function of the electric field and temperature \cite{Grabecki:1984_SS}. My assignment of some strange data to inversion layers at grain boundary defects  in p-Hg$_{1-x}$Mn$_x$Te \cite{Grabecki:1984_APL} allowed Grzegorz to show that the anomalous Hall effect, expected in the presence of magnetic ions, does not affect the precision of Hall resistance quantization \cite{Grabecki:1987_Pr,Grabecki:1993_SST}. At the same time,  Marian Herman (1936-2015) \cite{Herman:1989_B} and Janusz Sadowski were constructing an MBE setup for IV-VI compounds.

In 1987, Gorbachev decided to open boarders to Poland, closed in 1981. I immediately went to Moscow and Leningrad, where over an aromatic Armenian-style coffee in a tiny apartment inhabited by the Boris Altshuler family, my appetite to study mesoscopic phenomena in DMS spin-glasses grew even further.  We did not continue that discussion when Boris revisited me during Warsaw's ICPS a year later, as without any fear he used the time to ask other guests, Eva Andrei among them, where would it be better to work, in Europe or in the US? Unexpected by this, taking advantage of a relatively long mean free path of electrons in narrow-gap semiconductors we started to study mesoscopic universal conductance fluctuations (UCFs) just employing photolithography. That time Andre Geim entered to my office and seeing what was going on immediately advised us to use oil from not too good pumps in our scanning electron microscope as a resist. He actually appeared faster than we \cite{Grabecki:1991_APPA}, and reported already in 1990 mesoscopic spin-dependent UCFs as a function of the magnetic field in a GaAs  nanowire. However, I argued somewhat later that Andre and coworkers observed actually SdH oscillations, not UCFs \cite{Jaroszynski:1995_PRL}. Hearing complains coming from the mK lab that  UCFs in our n-Cd$_{0.99}$Mn$_{0.01}$Te:In submicron wire show an atypical temperature dependence, I realized that UCFs in paramagnetic DMSs originate primarily from the redistribution of carriers between spin subbands, whereas both the Aharonov-Bohm-type of interferences and spin-disorder scattering are of lesser importance \cite{Jaroszynski:1995_PRL}.  It took another year or two before  Ja\'s Jaroszy\'nski took a wonderful set of data for Cd$_{1-x}$Mn$_{x}$Te:I submicron wires in the spin-glass regime \cite{Jaroszynski:1998_PRL}. His analysis of the second spectra of conductance noise indicated that the Huse-Fisher droplet model applies to spin-glass freezing in random antiferromagnets.

There have been about 600 papers reporting results obtained on II-VI and later IV-VI quantum structures, often containing Mn, grown by IFPAN's MBEs. The results have constituted topics of invited talks at International Conferences on II-VI Compounds, and of the Tomasz Wojtowicz invited talk at the 34th ICPS, Montpellier 2018 and plenary talk at
the 18th International Conference on MBE, Flagstaff 2014.  Some of the findings are reviewed in the book edited by Jacek Kossut and Jan Gaj \cite{Kossut:2010_B} and in the book chapter co-authored by Aleksandr Kazakov and Tomasz Wojtowicz \cite{Kazakov:2020_B}. One example of numerous activity lines is the mesoscopic physics in submicron wires pattered of n-type doped DMS epilayers, as described above. Another research direction was exploring the suitability of self-organized quantum dots containing a single Mn ion as a quantum information carrier \cite{Goryca:2009_PRL}. Also, nanowires showing giant exchange splittings of excitonic lines were obtained \cite{Wojnar:2012_NL}.

An important research topic has concerned various quantum Hall effects (QHEs) in modulation-doped DMS quantum wells.  Due to large spin-splitting, the QHE data show a regular pattern, allowing for an accurate testing of the scaling theory \cite{Jaroszynski:2000_PE}.  However, in samples with a specific Mn concentration, mysterious narrow peaks are superimposed on beautiful QHE data. I then proposed that the quantum Hall ferromagnet (QHF) state, driven by electron-electron exchange, is formed on crossings of Landau levels, imposed by the giant $s$-$d$ exchange spin splittings \cite{Jaroszynski:2002_PRL}. Furthermore, a long electron mean free-path and giant exchange spin-splitting allowed the observation of Landau-Zener transitions between spin sublevels in a helical magnetic field produced by Dy microstripes \cite{Betthausen:2012_S}. Tomasz Wojtowicz's insistence in improving structures' quality made that the modulation-doped CdTe \cite{Piot:2010_PRB} and later Cd$_{1-x}$Mn$_x$Te \cite{Betthausen:2014_PRB} quantum wells belong to an exclusive club of systems in which the fractional QHE has been observed. Those accomplishments constitute a starting point of MagTop struggle aiming at using domains of the QHF in both QHE and FQHE regimes as a platform hosting Majorana and, possibly, non-Abelian excitations, as proposed in papers co-authored by MagTop researchers \cite{Kazakov:2017_PRL,Simion:2018_PRB}.

The  acquired proficiency in ultra high-vacuum techniques was a good starting point to initiate  partner collaboration between IFPAN and the highly successful PREVAC company, located in south-west Poland, and specializing in fabricating custom-designed ultra-high vacuum systems, such as MBE, ARPES, XPS, and others.  This collaboration is continued and reinforced by MagTop, as PREVAC is one of MagTop's industrial partners.

\section{The saga of dilute ferromagnetic semiconductors}
\label{sec:saga}

In November 2020, my wife Maniela got an sms from her nephew congratulating us that her support and my instance made me the most cited scientist working now in Poland.  The ensemble of papers on dilute {\em ferromagnetic} semiconductors  contributed mostly to such a position in the ranking completed according to a methodology elaborated by Stanford/SciTech Strategies/Elsevier bibliometric experts. (Surely, other methodologies would reshuffle the order). Also the EPS Condensed Matter Division Europhysics Prize, I received with David Awschalom and Hideo Ohno in 2005, and the 2006 Foundation for Polish Science Prize were for work on ferromagnetic semiconductors and semiconductor spintronics. On those occasions and later at similar ceremonial events, it has been underlined that our work bridged broadly and fundamentally the two leading fields of condensed matter physics: semiconductor physics and magnetism. We made mainstream nonmagnetic semiconductors ferromagnetic, in which the virtues of magnetic, electronic, and photonic systems are combined. This intersection of magnetism, semiconductor physics, materials science, and localization physics has led to discovery of new phenomena essential for new generations of spintronic devices, the initiation of a massive world-wide materials science search for magnetic semiconductors that resulted in the discovery of FeAs-based superconductors and variety of materials with ferromagnetic signatures that remain to be understood. Ferromagnetism in semiconductors has been essential in exploring new ideas and concepts, some of which -- like spin injection from a ferromagnet,  electrical magnetization manipulation, tunneling anisotropic MR, and spin-orbit torque -- have already been transferred to ferromagnetic structures of metals or complex oxides, others -- like interplay of spin-orbit coupling and exchange spin splitting -- to antiferromagnetic spintronics and research on topological matter. Accordingly, broadly understood spintronics has become one of the central themes of contemporary condensed matter physics and its ICT applications.

Two recent reviews describe findings in the field of dilute ferromagnetic semiconductors in more technical terms and in the context of word-wide research \cite{Dietl:2014_RMP,Dietl:2015_RMP}.  IFPAN's main accomplishments, achieved in collaboration  with Grenoble, Linz, and Sendai teams,  but also with Athens, Nottingham, Wuerzburg, and other centers, include:

\ \\
(1) {\em Predicting $T_{\text{C}}$s.} To establish conditions allowing for carriers'-mediated ferromagnetic ordering and to evaluate the magnitudes of Curie temperature $T_{\text{C}}$  and of various thermodynamic quantities, I made use of all that broad knowledge gathered over decades for the mainstream semiconductor compounds, i.e., III-V and II-VI tetrahedrally bonded semiconductors with carriers residing at the Briilouin zone center \cite{Dietl:1997_PRB,Dietl:2000_S,Dietl:2001_PRB}. While in Grenoble (altogether 7 months in 1994-1997), stimulated by Yves Merle d'Aubign\'e (1933-2000), I demonstrated the equivalence of the RKKY and Zener models, and predicted $T_{\text{C}}$s for various dimensionality structures of II-VI compounds in the mean-field approximation \cite{Dietl:1997_PRB,Dietl:1999_MSEB}. Zener's model [Phys. Rev. B 81, 440 (1951)] in this context means competition between spin entropy and carrier energy lowering coming from exchange splitting of the band).  The theoretically expected  $T_{\text{C}}$s  values were soon confirmed experimentally, particularly in the case of modulation doped p-Cd$_{1-x}$Mn$_x$Te:N quantum wells \cite{Haury:1997_PRL,Boukari:2002_PRL} (we understood the lack of hysteresis much later \cite{Kechrakos:2005_PRL,Lipinska:2009_PRB}), p-Zn$_{1-x}$Mn$_x$Te:N epilayers brought to IFPAN in 1998 for magnetic and transport characterization by their developers, then Ph.D. students, David Ferrand and Alberta Bonanni \cite{Ferrand:2000_JCG,Ferrand:2001_PRB}, and in IFPAN's bulk p-Zn$_{1-x}$Mn$_x$Te:N \cite{Khoi:2002_pssb,Kepa:2003_PRL}. Notably, the predicted absence of ferromagnetism above 1\,K in n-type samples was corroborated by millikelvin studies of n-Zn$_{1-x}$Mn$_x$O:Al \cite{Andrearczyk:2001_Pr}. Actually, much more impacting was my evaluation of $T_{\text{C}}$s, in the mean-field approximation and within the $p$-$d$ Zener model, for a broad range of p$^+$-type tetrahedrally coordinated semiconductors: the theory was in agreement with existing data (notably for Ga$_{1-x}$Mn$_x$As) and implied a survival of ferromagnetism to above room temperature in gallium nitride, zinc oxide, and diamond containing 5\% of Mn and $3.5\cdot10^{20}$ valence band holes per cm$^3$ \cite{Dietl:2000_S,Dietl:2001_PRB}. So far, no such itinerant hole densities were achieved in the latter systems, but the theory was quantitatively verified for  Ga$_{1-x}$Mn$_x$As, Ga$_{1-x}$Mn$_x$P, In$_{1-x}$Mn$_x$As, Ga$_{1-x}$Mn$_x$Sb, and In$_{1-x}$Mn$_x$Sb \cite{Dietl:2010_NM}, in which Mn acceptor radius is large enough to allow for hole delocalization. Also, pressure studies are consistent with the model \cite{Csontos:2005_NM,Gryglas:2020_PRB}. Other thermodynamic properties, such magnetization $M(T,H)$ and specific heat are also well described by the $p$-$d$ Zener model, provided that corrections to the mean-field theory are included \cite{Werpachowska:2010_PRB,Sliwa:2011_PRB}.

To put the understanding of carrier-mediated ferromagnetism in semiconductors in a broader context, it is worth contrasting it to another approach, in which the understanding often just means the identification of an {\em ab initio} code or an exchange-correlation potential that provides the magnetic groundstate observed experimentally.

\ \\
(2) {\em Micromagnetic properties}. I think it is hard to overvalue the beauty and power of micromagnetic theory which, together with the Landau-Lifshitz-Gilbert equation, describe magnificent magnetization patterns and their dynamics in ever more striking magnetic materials. But what about  material input parameters with which simulations are run? Typically, we rely on experimental determination, as only recently {\em ab initio} computations can provide an estimate of the magnetic anisotropy energy (MAE) of, say, bulk iron or a dilute ferromagnetic semiconductor Ge$_{1-x}$Mn$_x$Te \cite{Lusakowski:2015_JPC}. In contrast, our $p$-$d$ Zener model, built with the spin-orbit interaction taken into account, provided MAE magnitudes as a function of strain, temperature, hole and Mn concentrations in Ga$_{1-x}$Mn$_x$As \cite{Dietl:2000_S,Dietl:2001_PRB}. Combining that with the exchange stiffness theory developed by Allan McDonald's group, we were able, for instance, to describe stripe domains in Ga$_{1-x}$Mn$_x$As films with perpendicular magnetic anisotropy \cite{Dietl:2001_PRBb}. Furthermore, the predicted theoretically change in MAE sign as a function of carrier density \cite{Dietl:2001_PRB} was confirmed experimentally and explained in terms of the carrier redistribution between hole subbands \cite{Sawicki:2004_PRB}. Of course, specific materials science issues, such anisotropic spinodal decomposition (see, Sec.~\ref{sec:killed}) made the physics of MAE much richer \cite{Sawicki:2005_PRB,Wang:2005_PRL} than could be anticipated. Also, the presence of a superparamagnetic component, brought about by the proximity to the MIT, plays a role \cite{Sawicki:2010_NP,Sawicki:2018_PRB}. Results of micromagnetic studies of Ga$_{1-x}$Mn$_x$As and related ferromagnets by dozen groups worldwide (including our own \cite{Stefanowicz:2010_PRBa,Sawicki:2018_PRB}) show how far we have arrived with the understanding of these systems.

\ \\
(3) {\em Functionalities}. No doubt, the prospect of applications has been driving research in semiconductor spintronics. It was clear to me that the demonstration of changing the magnetic phase by light in modulation-doped p-Cd$_{1-x}$Mn$_x$Te quantum well told that other means, such as gating, would also provide a high degree of control over magnetic properties \cite{Haury:1997_PRL}, a dream made real in the $pin$ diode configuration in 2002, i.e., five years later \cite{Boukari:2002_PRL}. In the mean time, a series of ingenious works on magnetization manipulation by an electric field in gated In$_{1-x}$Mn$_x$As and Ga$_{1-x}$Mn$_x$As epilayers,  a major step towards energy efficient magnetization switching, carried out by Hideo Ohno and co-workers (with me as a discussion partner while at Tohoku) at the turn of the millennium, shook the magnetic and semiconductors communities. A contribution of IFPAN's researchers to this line of research was twofold. First, Maciej Sawicki constructed in Sendai and fully exploited a SQUID setup allowing measuring directly gate-induced magnetization changes \cite{Chiba:2008_N,Sawicki:2010_NP}.  Second, I developed a theory describing $T_{\text{C}}$ for systems with non-uniform carrier density \cite{Sawicki:2010_NP}, which successfully described the dependence of $T_{\text{C}}$ on the gate voltage in a range of Ga$_{1-x}$Mn$_x$As MOS FETs \cite{Nishitani:2010_PRB}. At the same time, Cezary \'Sliwa showed \cite{Stefanowicz:2010_PRBa} that the $p$-$d$ Zener model can nicely explain changes in the easy axis direction generated by gating \cite{Chiba:2008_N}. Another pioneering work done at Tohoku was the demonstration of magnetic wall displacement by an electric current without a magnetic field's assistance, the domain velocity being accurately described theoretically with my contribution \cite{Yamanouchi:2006_PRL}. At the same time, in the frame of European projects, Piotr Sankowski {\em et al.} have been writing tight-binding codes for computing spin injection in Zener-Esaki diodes \cite{Dorpe:2005_PRB} and magnetoresitance and anisotropic magnetoresistance in magnetic tunnel junctions \cite{Sankowski:2007_PRB}.
\ \\

When at immigration kiosks on four continents, I quoted "conference physicist" as my profession in 2001-2005. My 66 invited talks over those five years included plenary lectures at EPS meetings (Budapest 2002, Prague 2004, Bern 2005),  EMRS (Warsaw 2004),  27th ICPS (Flagstaff AR 2004), half-plenary at JEMS (Dresden 2003), and invited talks at Joint MMM Inter-Mag Conference (San Antonio TX 2001) and IUPAP's International Conference of Magnetism (Rome 2003). Similar visibility of David Awschalom and Hideo Ohno, together with influential writings presenting accomplishments, but also open questions (my contribution included Refs.\,\cite{Dietl:2001_APPA} and \cite{Dietl:2002_SST}), resulted in a widespread of spintronic research as well as in support by high-profile projects on semiconductor spintronics. IFPAN was part of the Ohno ERATO funding and E.\,C. collaborative FENIKS, AMORE, SPINOSA, NANOSPIN, and SemiSpinNet projects, as well as of my ERC FunDMS Advanced Grant.  This external funding was toped by the substantial support of government agencies and the non-public Foundation for Polish Science. Tomasz Wojtowicz, while with Jacek Furdyna at Notre Dame, contributed significantly to the DARPA project \cite{Ruster:2003_PRL}. More recently, the EAgLE project (2013-2016), obtained in the frame of the E.\,C. REGPOT initiative supporting long-term visits, helped lift the window to antiferromagntic spintronics \cite{Grzybowski:2017_PRL,Grzybowski:2019_AIP}. This project supported also my travels to present a half plenary talk at the ICM (Barcelona 2015) and an invited talk at the 33th ICPS (Beijing 2016), where I presented my view on the origin of spin-spin interactions in ferromagnetic topological semiconductors.

In this two-decades-long saga, incredibly unique was my one-year stay at Tohoku in 1999.  There, logistics was entirely ensured by Maniela, I had no meetings and virtually no teaching, all administration duties were taken over by Fumihiro Matsukura, and I could enjoy inspiring business lunches with Hideo Ohno in a tinny French-style restaurant, sometimes with Sukekatsu Ushioda, the future IUPAP President, at the table. Under these unique conditions and having a strong background in the DMSs physics \cite{Dietl:1994_HB} and in the envelope function formalism \cite{Szymanska:1978_JPCS,Dietl:1978_JPCS}, I was in position to identify properly (i) a minimal Hamiltonian suitable for describing carrier-mediated ferromagnetism in tetrahedrally coordinated p-type semiconductors; (ii) thermodynamic and micromagnetic quantities that theory can provide; (iii) relevant theory constrains, such as the metal-to-insulator transition, strong coupling, self-compensation, and solubility limits \cite{Dietl:2000_S}. It appears that over that year I was able to describe theoretically and examine numerically a more comprehensive set of phenomena and materials \cite{Dietl:2001_PRB} than a group of five brilliant postdocs (now professors across Europe) supervised by Allan MacDonald in Austin and Lu Sham in San Diego. Many papers' well-known bane is the lack of values of the parameters used, making it impossible to check the results. Surprisingly, one of the referees of our PRB \cite{Dietl:2001_PRB} recommended the removal of information on materials parameters I employed, apparently to make our work less substantial.

And at the end, a little too didactic illustration of the German proverb: {\em \"Ubung macht den Meister}.
I got terrific  Grenoble spectroscopic results on 2D ferromagnetism in p-Cd$_{1-x}$Mn$_x$Te quantum wells from Yves Merle d'Aubign\'e in February 1997, while at a family ski-week in G\"unther Bauer's secondary house in Bad Ischl.  Being busy with Warsaw's College of Science I could stay in Grenoble only one week in April. During that week, I managed to write down two manuscripts, this time without any help of IFPAN's English writing guru, Hanka Przybyli\'nska: (i) the PRL on the observation of 2D carrier-mediated ferromagnetism \cite{Haury:1997_PRL}, confirming nicely my a year old theoretical prediction \cite{Dietl:1997_PRB}, and (ii) the Rapid Comm on controlling excitonic reflectivity by giant exchange spin splitting in DMS photonic superlattices \cite{Sadowski:1997_PRB}, an extension of earlier Grenoble work, with my theoretical input, on non-magnetic photonic Bragg superlattices \cite{Merle:1996_PRB}. On returning to Warsaw I stopped Ewa Skrzypczak (1929-2020) on the Ho\.za street to admit that she was absolutely right one summer day twenty-eight years ago.  That day, she, as a deputy dean, consented to repeat exams in September.  Despite a crowd in front of her door, seeing my top grades in math and physics, and failed the lowest level English test, she decided to deliver to me a quarter of an hour tutorial on the significance of English proficiency in physics (though she might not know that I had similar failures with French and Russian while in Pozna\'n's Marcinek high school).

\section{Why the $p$-$d$ Zener model and superexchange, but neither double exchange nor Van Vleck mechanism?}
\label{sec:Zener}
I do not recall who approached me over coffee after a conference talk and friendly advised: "take it easy, do not treat that personally, be actually proud of it -- for many people, the way to magnify their visibility is to challenge the principal paradigm of the time". Indeed, soon after the appearance of our 2000 {\em Science} paper \cite{Dietl:2000_S}, it has become fashionable to contest its two crucial and mutually related presumptions: it has been argued that (i) the holes in Ga$_{1-x}$Mn$_x$As reside in an impurity band, not in the valence band; (ii)  the ferromagnetism of this material should be described in terms of double exchange, not by the $p$-$d$ Zener model. This situation motivated Tom\'a{\v{s}} Jungwirth at Prague's IoP CzAS to assemble arguments {\em against} the paradigm shift, and publish them with quite an impressive co-author list  \cite{Jungwirth:2007_PRB}.

As reviewed elsewhere \cite{Dietl:2014_RMP}, the trouble of models involving the presence of an impurity band is that all experiments designed to demonstrate directly its existence, especially photoemission and scanning tunneling spectroscopy have failed in Ga$_{1-x}$Mn$_x$As. Also, results of state-of-the-art {\em ab initio} band structure computations, such as hybrid functionals and the dynamic mean-field approximation, have not shown any impurity band. The data pointed out to a few percent admixture of Mn $d$ orbitals to the wave function of holes at the valence band top, just what is needed to explain the experimentally observed magnitudes of the valence band offset and the $p$-$d$ exchange energy. At the same time, the temperature dependence of low temperature conductivity, the Seebeck coefficient, infrared conductivity integrated over frequency implied the hole mass of the value similar to GaAs. Last but not least, as mentioned above, thermodynamic and micromagnetic properties of Ga$_{1-x}$Mn$_x$As, and also magnetic circular dichroism \cite{Dietl:2001_PRB}, can be described within the $p$-$d$ Zener model with no adjustable parameters -- no one has reported an attempt to do that within competing models.

All that is not to say that the physics and materials science of dilute ferromagnetic semiconductors is straightforward.  It was clear to me right from the beginning \cite{Dietl:2000_S} that the Anderson-Mott localization and the associated non-uniformities of magnetization, together with important issues of solubility limits and self-compensation as well as of the transition to a strong coupling case with decreasing the lattice constant need to be addressed experimentally. For instance, in the MIT vicinity, the resistivity tensor's absolute values can hardy be described quantitatively, and often even qualitative understanding is challenging \cite{Chiba:2010_PRL}. Furthermore, it is now known \cite{Dietl:2002_PRB,Dietl:2008_PRB}, also due to a beautiful series of works performed at UW by, among others, Agnieszka Wo{\l}o\'s and Marcin Zaj{\c{a}}c in the Maria Kami\'nska and Andrzej Twardowski groups, respectively, that GaN:Mn is in a strong coupling limit ($p$-$d$ hybridization alone can bind a hole on the Mn ion), a conclusion consistent with parallel studies carried at Schottky in Garching. This meant, disappointingly, that resulting mid-gap Mn acceptors would not introduce any valence band holes that might lead to room temperature ferromagnetism we had predicted \cite{Dietl:2000_S}. Interestingly, however, Marcin, after completing his Ph.D. degree, moved to Robert Dwili\'nski's AMMONO (the company that made  the IEEE Spectrum's cover story; July 2010 issue), and introduced Mn as a trap of residual donor electrons, always present in GaN. Now, 1.5" semiinsulating GaN:Mn substrates are the flagship product of the AMMONO/Unipress company.

To understand the Ga$_{1-x}$Mn$_x$N physics, especially fruitful has been collaboration with Alberta Bonanni's group at Kepler University in Linz, where Ga$_{1-x}$Mn$_x$N epilayers have been obtained by MOVPE.  More recently, Detlef Hommel, now in Wroc{\l}aw, has entered with his co-workers and MBE-grown films to the loop. Extensive characterization employing various photon, electron, and ion beams demonstrated world-top crystal quality, low electron concentration (measured in GaN), and random Mn distribution with no spinodal decomposition up to $x = 0.1$ \cite{Stefanowicz:2010_PRB,Bonanni:2011_PRB,Sawicki:2012_PRB,Kunert:2012_APL,Stefanowicz:2013_PRB}. Seeing low-temperature ferromagnetism of Mn$^{3+}$ ions in these samples, I asked Jacek Majewski to dig up his tight-binding code that had served a decade earlier to predict ferromagnetic superexchange coupling between Cr$^{2}$ ions in II-VI compounds \cite{Blinowski:1996_PRB}. Combining it with Monte Carlo simulations we obtained a successful description of $T_{\text{C}}(x)$ \cite{Sawicki:2012_PRB,Stefanowicz:2013_PRB}.

Why, however, despite that the Fermi energy is pinned by the Mn$^{2+}$/Mn$^{3+}$ impurity band, do we speak about superexchange, not about double-exchange invented also by Clarence Zener? The key reason is that double exchange requires certain delocalization of $d$ band carriers, the case of, for instance, CMR oxides. By contrast, in Ga$_{1-x}$Mn$_x$N, the dilution and the random Mn distribution together with a sizable Jahn-Teller effect and short localization radius in the strong coupling limit make that we are far from the MIT, deeply in the strongly localized regime. In agreement with the expectation of the Anderson-Goodenough-Kanamori superexchange theory, we observed Mn-Mn coupling to be ferromagnetic if a majority of Mn ions is in the 3$+$ configuration. In contrast, antiferromagnetic interactions take over if Mn$^{2+}$ prevail, the case of Ga$_{1-x}$Mn$_x$N samples with a high concentration of compensating donors. In particular, at least so far, we have not found any $T_{\text{C}}$ maximum at half filling, as expected within the double exchange scenario.  The semiinsulating character of Ga$_{1-x}$Mn$_x$N led me to suggest that the influence of an electric field on magnetic properties resulted from sample deformation by the inverse piezoelectric effect \cite{Sztenkiel:2016_NC}, not by an alternation of carrier density in the impurity band.

At the turn of the 1970s and 1980s, G\'erald Bastard, the future honorary chair of the 34th ICPS (Montpellier 2018),  worked with coworkers to understand magnetization behavior $M(T,H)$ in Mn-doped and Fe-doped HgTe.  They proposed, in particular, that {\em antiferromagnetic} coupling between Mn ions results from the Bloembergen-Rowland (BR) mechanism, an interband analog of the RKKY interaction, which could be strong in the zero-gap (topological) HgTe, the coupling also considered by me and others in Warsaw \cite{Ginter:1979_pssb}. We now know that antiferromagnetic superexchange actually dominates \cite{Kacman:2001_SST} (see, Sec.~\ref{sec:DMS}). In contrast, in the light of later studies of various Fe-based DMSs in several labs, the assignment of a linear field-dependence $M(H)$ to the Van Vleck paramagnetism was correct (Serre {\em et al.}, Proceedings of the Linz Narrow-Gap Conference, Linz 1981).  In the ground-breaking theory paper of the IoPChAS/Stanford/Tsinghua collaboration on the quantization of the anomalous Hall resistance in ferromagnetic topological materials (Science 2010), the authors used the formula I derived to study the BR contribution in p-Ga$_{1-x}$Mn$_{x}$As \cite{Dietl:2001_PRB} to claim that the BR mechanism, called by them the Van Vleck paramagnetism, accounts for {\em ferromagnetic} coupling between Cr spins in topological Bi$_2$Se$_3$.  I contend, looking on Czarek \'Sliwa results for HgTe \cite{Sliwa:2021_PRB} as well as on {\em ab initio} computations and experimental data for various transition metals in tetradymite semiconductor compounds that -- leaving aside the interaction name -- also in topological Cr- and V-doped Bi and Sb chalcogenides, not the BR mechanism but superexchange prevails, but of the ferromagnetic sign, just as in the case of Mn$^{3+}$ in GaN. Actually, as we have shown \cite{Sliwa:2021_PRB}, the Van Vleck story is quite instructive: the ferromagnetism came about by an implicit inclusion of the self-interaction term $i = j$, instead of summing up only over TM pairs $i \ne j$.

\section{How I killed semiconductor spintronics and why it had it coming}
\label{sec:killed}
Of course, this section title is a paraphrase of "How I Killed Pluto and Why It Had It Coming" - a marvelous book by Mike Brown. In 2003, I started to worry that our prediction about the room temperature ferromagnetism in wide-band gap DMSs had been becoming too ... {\em successful} \cite{Dietl:2003_NM}: high $T_{\text{C}}$ was reported for DMSs, in which no ferromagnetism was expected or even, one year later, for compounds nominally without any magnetic ions! Moreover, Maciej Sawicki's magnetic data \cite{Karczewski:2003_JSNM} and some other group results pointed to the presence of magnetic precipitation. Thus, the crucial challenge became proper nanocharacterization, a demanding task, particularly considering that I \cite{Dietl:2005_Pr} and Hiroshi Katayama-Yoshida's group in Osaka realized that precipitates may assume the host crystal structure, driven by chemical spinodal nanodecomposition.

To go to the point: the advanced nanocharacterization protocols developed by Alberta Bonanni in Linz for (Ga,Fe)N \cite{Bonanni:2007_PRB,Bonanni:2008_PRL},  Shinji Kuroda in Tsukuba for (Zn,Cr)Te \cite{Kuroda:2007_NM}, and by others, together with results of SQUID magnetometry  (that has reached the art level \cite{Sawicki:2011_SST,Gas:2021_JAC}) and progress in using appropriate {\em ab initio} tools, achieved in Osaka, Golden Colorado, and other places shed new light on the interplay of DMS materials science and magnetic properties \cite{Dietl:2008_JAP,Bonanni:2010_CSR,Dietl:2015_RMP}:

\ \\
(1) {\em Origin of high $T_{\text{C}}$s}. Ferromagnetic-like features surviving up to above room temperature originate from the presence of nanoregions with a large density of the magnetic constituent and, thus, characterized by a high spin ordering temperature.  The interplay of attractive chemical forces between magnetic ions (driven by a contribution of the $d$ orbitals to bonding), entropy, and kinetic barriers determine the degree of non-uniformity in the magnetic ion distribution, i.e., an effective solubility limit, for given growth conditions and post-growth processing. The resulting spinodal nanodecomposition can have a character of chemical or crystallographic phase separation, in the latter case precipitation of metal nonoparticles sometimes occur. Unforseen contamination by magnetic nanoparticles can also be involved in specific cases. Uncompensated spins on surfaces of antiferromagnetic nanocrystals may contribute to magnetic response \cite{Dietl:2007_PRB}. Furthermore, various coupling kinds  between magnetized regions increase apparent superparamagnetic blocking temperature \cite{Sawicki:2013_PRB}. In general, however, magnetic hystereses are tilted with a minute, if any, remanent magnetization and coercivity. No clear correlation exists between $T_{\text{C}}$ and the nominal concentration of magnetic impurities.

\ \\
(2) {\em Controlling aggregation of magnetic ions}. As demonstrated for (Zn,Cr)Te \cite{Kuroda:2007_NM} and (Ga,Fe)N \cite{Bonanni:2008_PRL}, the aggregation efficacy and, thus magnetic ordering, can be controlled by co-doping with shallow impurities that changes the charge state of magnetic ions and, thus, chemical forces between them \cite{Dietl:2006_NM}. In some cases, impurity complexes containing magnetic ions and shallow dopants appear, the case of (Ga,Mn)N doped with Mg, in which, as demonstrated by the Linz/Warsaw collaboration, clusters of Mn with one to three charged Mg acceptors bound to it account for magnetic and optical properties of Mn ions \cite{Devillers:2012_SR}. A somewhat related is a self-compensation effect predicted for Ga$_{1-x}$Mn$_x$As by Jan Ma{\v{s}}ek and Frantisek Mac\`a in Prague, and found experimentally by Berkeley/Notre Dame researchers, who unsuccessfully tried to raise $T_{\text{C}}$ by increasing the concentration of Mn or Be acceptors. Their glorious Rutherford backscattering data demonstrated that to halt an energy increase associated with a decrease in density of electrons participating in bonding (i.e., hole formation), Mn ions displace to interstitial donor positions. In this location, they lower $T_{\text{C}}$, as they couple antiferromagnetically to substitutional Mn ions \cite{Blinowski:2003_PRB}. A process of interstitials' removal by low-temperature annealing, which increases $T_{\text{C}}$, was then quantitatively examined \cite{Edmonds:2004_PRL}.

\ \\
(3) {\em  Anisotropic spinodal decomposition}. At some point it became clear to me that conditions at the growth front are essential for the resulting distribution of magnetic ions. For instance, the fact that chemical force was found attractive for Fe ions but repulsive in the case of Mn pairs on the GaN surface \cite{Gonzalez:2011_PRB} explained why it was possible to grow Ga$_{1-x}$Mn$_x$N epilayers with a random distribution of Mn ions. Another important consequence of this insight was the realization that the magnitude of attractive force at the surface is larger for the nearest neighbor Mn cation dimer residing along $\langle 1{\bar{1}}0\rangle$ direction compared to the $\langle 110\rangle$ arrangement, as there is no anion bonding the  $\langle 1{\bar{1}}0\rangle$ pair on the (001) surface of the zinc-blende crystals \cite{Birowska:2012_PRL}. A beautiful group theoretical analysis of crystals with  $\langle1{\bar{1}}0\rangle$ dimers was carried out by Czarek \'Sliwa. His evaluations showed that indeed the anisotropic spinodal decomposition can explain the origin and the magnitude of the in-plane uniaxial magnetic anisotropy hitherto mysterious but essential for functionalities of Ga$_{1-x}$Mn$_x$As \cite{Birowska:2012_PRL}. This symmetry breaking, i.e., nematicity, is particularly strong in In$_{1-x}$Fe$_x$As and was directly revealed by element sensitive nanocharacterization \cite{Yuan:2018_PRM}. Since this discovery, I have been promoting the idea that the nematicity -- observed, for instance, in the quantum Hall effect, unconventional superconductors, and magnetic oxides -- originates in many cases from quenched anisotropic distribution of defects, impurities, or alloy components formed during the growth rather than from  a spontaneous symmetry breaking in the many-body electronic subsystem cooled down to sufficiently low temperatures.
\ \\

It has been more than two thousand papers claiming the discovery of high temperature ferromagnetism in semiconductors, oxides, various forms of carbon, topological matter, and most recently, in atomically thin 2D layers containing a low concentration of magnetic constituent (say, $x <10$\%) or nominal without any magnetic ions, $x =0$. Considering the lack of follow up reports on spintronic functionalities at room temperature, it appears right to agree with Alex Zunger, who once remarked: "those who carefully characterize their samples and worry about the reproducibility deprive themselves of publications in high impact journals". However, taking into account foreseen functionalities of metal/semiconductor/magnet hybrids and prospects of self-organized growth of novel magnetic nanostructures \cite{Dietl:2008_JAP,Dietl:2015_RMP} as well as steady progress in controlling properties of embedded magnetic nanocrystals \cite{Navarro:2020_M}, it is quite possible that studies of magnetic nanocomposites may return to many labs. And somewhat related question: do we need high temperature uniform ferromagnetic or antiferromagnetic semiconductors for future spintronic devices? As mentioned in the previous section, spintronic functionalities discovered in dilute ferromagnetic semiconductors operate wonderfully at room temperature in ferromagnetic metals. We should also not forget that ferrite antennas of ferrimagnetic oxides have been with us since many decades. Nevertheless, similarly to the case of superconductors, we will not stop struggling to push operating temperature higher of, say, the quantum anomalous Hall effect in samples of topological ferromagnetic or antiferromagnetic semiconductors \cite{Pournaghavi:2021_PRB}, considering foreseen applications in quantum metrology and possibly in topological quantum computing.

\section{And yet the impurity band -- the trio of quantum Hall effects}
I have already mentioned in Sec.~\ref{sec:resonant} that the presence of the acceptor impurity band accounts for surprisingly high values of low-temperature electron and hole mobilities at the 3D and 2D topological phase transitions, respectively, in HgTe-based systems. Interestingly, this band plays an
essential role in the quantum Hall effect (QHE), quantum spin Hall effect (QSHE), and quantum anomalous Hall effect (QAHE).

Since high-school times, we know that strong response to electrostatic gating and doping with carriers'- providing impurities accounts for the widespread use of semiconductors in information communication chips and new energy technologies, such as solar cells. These charge dopants, vital for applications, can be intentionally introduced by materials growers or originate from lattice defects or foreign background atoms always present in real materials. A virtually exact Hall resistance quantization to a value reliant only on the Planck constant $h$ and electron charge $e$ in imperfect samples came in 1980 as a major surprise. However, we now understand that it results from the lack of backscattering in edge channels in high magnetic fields, which leads to resistance quantization in the 1D case. According to this insight, the quantization precision deteriorates in narrow samples \cite{Wrobel:1998_PRB}, in which scattering between channels adjacent to opposite edges of the Hall bar is possible in real devices.

Actually, a closer look demonstrates that background dopants are not an obstacle but a resource in the QHE: by pinning the Fermi energy in the region between the centers of Landau levels hosting an extended state, the localized states brought about by impurities and defect potentials, account for a non-zero width of quantized Hall resistance plateaus in 2D systems. The presence of sufficiently wide plateaus is essential for metrological applications of the QHE, and played a crucial role in redefining S.I. units in terms of constants of nature in 2019. I argue that a similar situation takes place in the case of the QSHE and the QAHE, discovered experimentally by Laurens Molenkamp's group in Wuerzburg and  Qi-Kun Xue's team in Beijing, respectively, where the pinning of the Fermi level by a localized impurity band (residing in the energy gap) allows for the observation of carrier transport by topological edge states in a specific range of gate voltages. However, there is a significant difference between those two phenomena: in the case of QSHE, two time-reversal partners (helical states) form edge conducting channels, whereas time-reversal breaking by ferromagnetically ordered Cr or V spins results in the presence of a single chiral edge channel in the QAHE regime.

Motivated by experimental results obtained on Wuerzburg's and Novosybirsk's samples containing HgTe quantum wells, the latter brought to Warsaw by Wojtek Knap, and investigated comprehensively by Grzegorz Grabecki and Magdalena Majewicz at IFPAN and Ivan Yahniuk at Unipress \cite{Grabecki:2013_PRB,Majewicz:2014_APPA,Yahniuk:2019_QM,Yahniuk:2021_arXiv},  I considered the roles of background acceptors in QSHE materials in two companion papers \cite{Dietl:2023_PRL,Dietl:2023_PRB}. Importantly, the concentration of those dopants can be evaluated from the gate-voltage width corresponding to the bandgap region.

The theory's starting point is the determination of acceptor ground-state energies depending on the location of the dopant in respect to the quantum well center. In this way, we obtain information about the location of the acceptor impurity band in respect to valence and conduction  bands in the QW.  Importantly, donor impurities does not give localized states in this system. The next step has provided the magnitude of the exchange interaction between edge electrons and acceptor holes, coupled by the Coulomb force and $kp$ interaction, determined making use of the previously obtained information on the penetration of the edge channels into the quantum well \cite{Papaj:2016_PRB}. The role of anisotropic $s$--$p$ and $sp$--$d$ exchange interactions, Kondo coupling, Luttinger liquid effects, precessional dephasing, and bound magnetic polarons is quantified for HgTe and (Hg,Mn)Te QWs. I found that the resulting scattering rate between edge channels explained the experimentally found length electron travels without backscattering in topological HgTe quantum wells and WTe$_2$ monolayers, typically of the order of 10 and 0.1 $\mu$m, respectively. While these values are in agreement with experimental data, they are several orders of magnitude shorter than in the case of QHE and QAHE. Despite constraints associated with momentum and angular momentum conservation, the backscattering rate is relatively high due to a sizable magnitude of electron-hole exchange, leading to the strong coupling limit of the Kondo effect, where the spin-flip scattering attains the unitary limit.

Interestingly, Fermi-level pinning by negative-$U$ centers with two bound carriers in a singlet state or doping with paramagnetic impurities like Mn may improve quantization precision. The latter looks surprising, but quantitative evaluation demonstrates that rather than showing the Kondo effect, Mn spins form ferromagnetic clouds around acceptors (i.e., BMPs), diminishing Kondo temperature of dopants and, thus, reducing the backscattering rate at low temperatures \cite{Dietl:2023_PRB}, as observed recently in Wuerzburg.

\section{On the road to topological superconductivity}

It was a wonderful night. It happened to me during a carnival period in 1987. Suddenly, when sweeping the horizontal magnetic field at 30\,mK,  the $xy$ recorder's needle started moving down, and after a dozen of seconds up, reviling a strong diamagnetic signal from our home-made a.c. magnetometer \cite{Lenard:1994_C}. That diamagnetism appeared when the Earth magnetic field was compensated by the field generated by a Helmholz coil I had recommended to install around the millikevin cryostat. This finding confirmed the origin of the resistance drop below 0.5\,K, found by Maciej Sawicki during his night shift a couple of weeks earlier, once he had decided to decrease the current to an irrationally low value. I still remember vibrating silence that fell in the Jaszowiec 1987 lecture hall when at the end of my talk "Physics of semiconductors below 1\,K" \cite{Dietl:1988_APPA}, after describing quantum localization phenomena in Mn-based DMSs and the latest literature results on the fractional quantum Hall effect,  I disclosed our discovery of superconductivity in zinc-blende Bridgman grown Hg$_{1-x}$Fe$_x$Se crystals.

A detailed analysis showed that untypically low magnitudes of critical currents and fields as well as specific dependencies of resistivity and magnetic susceptibility on temperature can rather nicely be explained by the presence of few superconducting precipitates and the proximity effect \cite{Lenard:1990_JLTP}. Actually,  unusually high electron mobility in HgSe:Fe, described in Sec.\,\ref{sec:resonant}, accounted for a considerable Cooper pair diffusion length in HgSe:Fe. Using electron microscopy we revealed indeed the presence of some inclusions but we were unable to determine their nature. After annealing in Se vapor,  the disappearance of superconductivity suggested that Hg droplets might be involved \cite{Lenard:1990_JLTP}. I consider not asking Andrzej Mycielski that time to grow FeSe, as the greatest shame in my scientific career.

In 1986 superconductivity features were found in PbTe-SnTe superlattices by Akihiro Ishida and co-workers at Shizuoka University and, over the next two decades, in a broad range of topological and non-topological IV-VI superlattices and heterostructurs grown by  Alexander Sipatov and co-workers in Kharkov since 1988. Prompted by these results, Evelyne Tang and Liang Fu at MIT proposed in 2014 a theory of superconductivity, whose essential ingredient was a flat band formed by a periodic arrangement of misfit dislocations in topological materials. In our lab, we found superconductivity in Kiev's bulk PbTe \cite{Darchuk:1998_Semicon} and in IFPAN's bulk Pb$_{0.63}$Sn$_{0.37}$Se \cite{Mazur:2019_PRB}, presenting in both cases strong arguments that metallic Pb and Sn precipitates, respectively were a source of Cooper pairs. Somewhat related and also surprising were Andreev reflection results for In/PbTe \cite{Grabecki:2010_JAP} and In/NbP heterostructures  \cite{Grabecki:2020_PRB}, taken by Grzegorz Grabecki {\em et al.} at IFPAN. The former pointed to $T_c$ magnitudes higher than that of metallic In. In the simplest interpretation, In diffusion into PbTe resulted in  superconducting Pb precipitates \cite{Grabecki:2010_JAP}. In contrast, the In/NbP data suggest the presence of a low-gap superconductivity, presumably appearing in the Weyl semimetal NbP, proximitized by the superconducting In \cite{Grabecki:2020_PRB}.

Even more striking is the observation by several groups of Andreev-like spectra in topological semiconductors and semimetals, nominally without any superconductor, i.e., by making point contacts of normal (e.g., Ag, Au) or even ferromagnetic (e.g., Co) metals. For instance, Grzegorz Mazur and Krzysztof Dybko, employing a brand new Triton 400 dry dilution refrigerator with a vector magnetic field, installed in our lab by Maciej Zgirski and Marek Foltyn, found such spectra in samples with silver paint contacts to diamagnetic Pb$_{1-y}$Sn$_{y}$Te as well as to paramagnetic or ferromagnetic Pb$_{1-x-y}$Mn$_x$Sn$_{y}$Te with Sn content $y$ corresponding to the topological crystalline insulator phase  \cite{Mazur:2019_PRB}. Is this phenomenon caused by hardly detectable residual precipitates of superconducting metals? Does interfacial topological superconductivity account for it, as suggested in virtually all publications? Could it originate from topological gap states at domain walls of a 1D carrier liquid at surface atomic steps, as proposed by Wojciech Brzezicki {\em et al.} \cite{Brzezicki:2019_PRB}? If yes, how to make the discovery useful for sensors and/or topological quantum computation? This is one set of questions addressed by researchers from 13 countries at the International Centre for Interfacing Magnetism and Superconductivity with Topological Matter MagTop, heading since 2017 by me and Tomasz Wojtowicz at IFPAN.

\section{Epilog}
In spring 2020, I was invited by UW young professors Krzysztof Pachucki and Piotr Wasylczyk to one of the on-line weekly discussion meetings they organize for a selected group of students. To start, I  showed them viewgraphs with pieces of advice on how to run in the front, not behind the others (of the sort that publishing is essential but more important is uncovering a novel challenge, and solving it). However, the main advice was to realize that the research landscape is dynamic: useful tips in one epoch, fails in another. For instance, I was fast in appreciating the role of arXiv, and  applying in the first call for the KBN project and, later, for the E.C, Maestro, AdG ERC, and IRA projects. In turn, I was too late in abandoning meaningless titles, such as "Millikelvin Studies of Mixed--Valence HgSe:Fe" and conference proceedings (but surely one should move around, to be inspired by great people and wonders of the world).

After consulting  Andrzej Mycielski, I started my pre-MSc experimental work in 1971 under the supervision of Andrzej J{\c{e}drzejczak, who using his sophisticated thermoelectric setup,  needed help in urgent collecting of the Nernst-Ettingshausen data for n-InSb, to be presented by Wlodek Zawadzki in his plenary talk at the 9th ICPS (Warsaw 1972). In a lecture delivered at a meeting of Student Research Groups in Physics in {\L}\'od\'z that fall, I described my fascination, lasting until now, that a piece of a semiconductor is a laboratory, in which we can observe and quantify the flow of electrons under electric and magnetic fields.  It was, however, a little later when I started to experience a passionate, to do not say sensual, pleasure at that moment when the intricate puzzle turned out into a clear picture –- it could be the realization that magnetic hysteresis at mK temperatures are generated by a piece of type II superconductor used for wires' soldering; it could be the recognition that low-field SdH oscillations in a p-type sample originate from a grain boundary in a nominally perfect single crystal; it could be the understanding that thermodynamic fluctuations or the inverse piezoelectric effect constitute the mechanisms explaining surprising optical or magnetization data. I think that such moments of illumination (to refer to Krzysztof Zanussi's film) have actually been the main driving force on my way.

\section*{Acknowledgments}
I would like to thank my wonderful family (now enlarged by nine grandchildren) for continuous support and encouragement, and my past and present collaborators for the great time we have had together. The International Research Centre MagTop is funded by the Foundation for Polish Science through the IRA Programme financed by EU within SG OP Programme.


%

\end{document}